\documentclass{elsart}
\usepackage{graphicx}
\usepackage{pxfonts}
\usepackage{lineno}
\usepackage{appendix}
\usepackage{amssymb}
\usepackage{footnote}%!!!
\usepackage{multirow}
\usepackage{varwidth}
\usepackage{array}
\usepackage{epstopdf}
\usepackage{url}

\journal{}

\begin{document}

\thispagestyle{empty}
\begin{Large}
	\textbf{DEUTSCHES ELEKTRONEN-SYNCHROTRON}
	
	\textbf{\large{Ein Forschungszentrum der Helmholtz-Gemeinschaft}\\}
\end{Large}

PUBDB 00223

January  2025

\begin{eqnarray}
\nonumber &&\cr \nonumber && \cr \nonumber &&\cr
\end{eqnarray}
\begin{eqnarray}
\nonumber
\end{eqnarray}
\begin{center}
	\begin{Large}
		\textbf{Concept of an Infrared FEL for the Chemical Dynamics Research Laboratory at PETRA IV}
	\end{Large}
	\begin{eqnarray}
	\nonumber &&\cr \nonumber && \cr
	\end{eqnarray}
	
	%	\begin{large}
	%		Gianluca Geloni,
	%	\end{large}
	%
	%	\textsl{\\European XFEL GmbH, Schenefeld}
	%	\begin{large}
	
	Evgeny Saldin
	%	\end{large}
	%
	\textsl{\\Deutsches Elektronen-Synchrotron DESY, Hamburg}
	\begin{eqnarray}
	\nonumber
	\end{eqnarray}
	\begin{eqnarray}
	\nonumber
	\end{eqnarray}

	\begin{eqnarray}
	\nonumber
	\end{eqnarray}
	\begin{large}
		\textbf{NOTKESTRASSE 85 - 22607 HAMBURG}
	\end{large}
\end{center}
%\end{widetext}
\clearpage
\newpage

\begin{frontmatter}

% Title, authors and addresses

% use the thanksref command within \title, \author or \address for footnotes;
% use the corauthref command within \author for corresponding author footnotes;
% use the ead command for the email address,
% and the form \ead[url] for the home page:
% \title{Title\thanksref{label1}}
% \thanks[label1]{}
% \author{Name\corauthref{cor1}\thanksref{label2}}
% \ead{email address}
% \ead[url]{home page}
% \thanks[label2]{}
% \corauth[cor1]{}
% \address{Address\thanksref{label3}}
% \thanks[label3]{}

\title{Concept of an Infrared FEL for the Chemical Dynamics Research Laboratory at PETRA IV}

% use optional labels to link authors explicitly to addresses:
% \author[label1,label2]{}
% \address[label1]{}
% \address[label2]{}

%\author[XFEL]{Gianluca Geloni,}
%\author[DESY]{Vitali Kocharyan,}
\author[DESY]{Evgeny Saldin}
\address[DESY]{Deutsches Elektronen-Synchrotron DESY, Hamburg, Germany}
%\address[XFEL]{European XFEL GmbH, Hamburg, Germany}

\begin{abstract}

We describe an infrared free electron laser (FEL), proposed as a part of a user facility that also incorporates synchrotron-radiation beamlines for the PETRA IV. The FEL itself addresses the needs of the chemical sciences community for a high-brightness, tunable source covering a broad region of the infrared spectrum - from 5 to 100 mkm. The user facility will allow, for the first time, the integrated and simultaneous use of dedicated infrared FEL and synchrotron-radiation beamlines for pump-probe experiments that will focus on gaining a rigorous molecular-level understanding of combustion and other energetic molecular processes. These (pump-probe) requirements dictate the use of storage ring RF structures and cw operation. 
The technical approach adopted in FEL  design uses an old PETRA III RF system and accelerating cavities. The primary motivation for adopting this approach was to minimize facility costs.

\end{abstract}

%\begin{keyword}
%
%% keywords here, in the form: keyword \sep keyword
%%edge radiation \sep near-field \sep electron-bunch diagnostics
%%\sep x-ray free-electron laser (XFEL)
%

%% PACS codes here, in the form: \PACS code \sep code
%%\PACS 41.60.Cr \sep 42.25.-p \sep 41.75.-Ht
%\end{keyword}
%
\end{frontmatter}

% main text

%\linenumbers

\section{ Introduction }

Since the end of 2009, the PETRA storage ring has been operated as part of the third-generation synchrotron radiation source PETRA III. With the conceptual design presented in \cite{CDR}, a next, fourth-generation synchrotron radiation source is proposed: PETRA IV. Further, free-electron lasers (FELs) have already operated successfully at infrared wavelengths \cite{N}. This paper describes such an FEL proposed as a part of the chemical dynamics research laboratory, a user facility that also incorporates beamlines for the PETRA IV. 
The FEL itself addresses the needs of the chemical sciences community for a high-brightness, tunable source covering a broad region of the infrared spectrum - from 5 to 100 mkm. 
The radiation sources can be synchronized to permit powerful two-color, pump-probe experiments that will further our fundamental understanding of chemical dynamics at the molecular level, especially those aspects relevant to practical issues in combustion chemistry.      

To minimize the influence of RF faults on the beam, the installation of a new RF system and accelerating cavities  at PETRA IV is planned.
The technical approach adopted in FEL  design makes use of the old PETRA III RF system and accelerating structures. The primary motivation for adopting this approach was to meet the user requirements for pump-probe experiments and to minimize facility costs.  The continuous wave (CW) mode of operation offers considerably higher average output power. It also allows for various pulse-gating configurations that will permit simultaneous multi-user operation.  

The proposed facility design is based on two turns of copper energy recovery linac (ERL). Multi-turn ERL looks very promising for making the facility less expensive.
A system at Novosibirsk  \cite{B} shows the degree to which CW copper machine performance can be extended. A multi-pass recirculation and energy recovery copper linac at Novosibirsk operates CW at the frequency 180 MHz. The accelerator facility consists of five acceleration and five deceleration passes. It comprises three FEL stages. The tuning range of these FELs is from 5 mkm to 240 mkm and the average power is about 1 kW and the peak power of 1 MW.

There have been active efforts in the US to design an infrared FEL for pump-probe experiments that incorporate synchrotron radiation beamlines for the LBL's  third-generation light source ALS \cite{LBL}. The project was not realized because of the costly superconducting accelerator infrastructure.
We have incorporated these ideas into a new room-temperature facility design that should lead to significant cost savings over the LBL design.

\section{ Scientific Opportunities}

The Europe Energy Strategy requires that combustors be designed and operated with unprecedented fuel efficiency and low levels of emitted pollutants. If these requirements are to be met, combusting chemistry must be understood at a much deeper level.  At the chemical dynamics research laboratory, the scientific research program will focus on the fundamental study of the chemistry of combustion.

The chemical research laboratory will allow, for the first time, the integrated  and simultaneous use of dedicated infrared FEL and synchrotron radiation beamlines for pump-probe experiments. One mode will use the PETRA IV photon beam as the pump and the FEL photon beam as the probe; in this mode, researchers will be able to study the vibrational structure of highly excited molecules. The other mode - FEL pump and PETRA IV probe  - will yield infrared spectra of transient species (combusting diagnostics); new information on ionization potentials and bond energies of transient species; and new vibration spectra of ions. 
It should be emphasized that infrared FEL offers an enormous range of research possibilities. Besides providing important opportunities for understanding the dynamics of combustion, the available resources will advance the broader field of chemical dynamics in general.

\section{Accelerating Facility Description}

The technical approach adopted in accelerator  design uses an old PETRA III RF system and accelerating cavities. 
The RF system of the PETRA III storage ring has been generating a voltage of 20 MV at a frequency 499.6 MHz and accelerating an electron beam with current of 100 mA for synchrotron radiation users. The required RF power is generated by klystrons in two RF stations and supplied two 12 seven-cells cavities in strait section. The PETRA III is driven by the RF power of two transmitters, each equipped with two 800 kW klystrons.   
When updating the storage ring for PETRA IV, the installation of 24 single-cell cavities is planned. Instead of klystron, solid-state amplifier transmitters are planned for for the PETRA IV RF system. 

\begin{figure}
	\centering
	\includegraphics[width=0.8\textwidth]{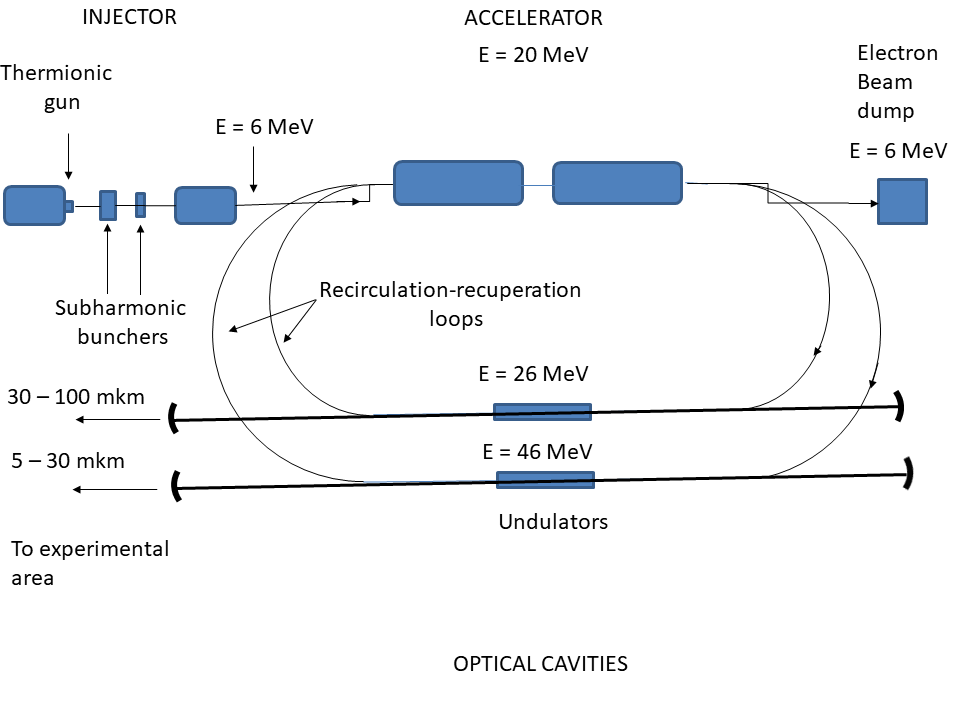}% \hfil% Here is how to import EPS art
	\caption{Schematic plan view of the infrared FEL facility.}
	\label{B103}
\end{figure}

Fig. \ref{B103} shows the layout of the RF accelerator system together with the injector. 
The electron beam must pass twice through the accelerator section to reach the beam energy greate than 26 MeV. A recirculation path thus routes the 26 MeV beam back to the entrance of the first RF cacity for further acceleration to energy up to 46 MeV. 
Fig. \ref{B103} shows the overall beam transport system. It consists of separate sections: 180$^0$ achromatic isochronous arcs, corrected to second order; Transport sections between RF cavities and the 180$^0$ arcs, and from 40 MeV 180$^0$ arc to the FEL; Transport lines to the beam dump.

The injector system is of critical importance to the overall performance of the FEL. The emittance, energy spread, current, and bunch length of the electron beam are all primarily determined by the performance of the injector. 
The design adopted uses a conventional thermionic cathode, subharmonic bunchers, and fundamental acceleration. This arrangement is shown schematically in Fig. \ref{B103}. The accelerator output characteristics are: Charge 2 nC, bunch length 30 ps, peak current 70 A, Normalized rms emittance 20 mm$\cdot$mrad, energy spread 0.5 percent at 46 MeV, bunch frequency 6.1 MHz, energy stability 0.005 percent. The gun bunch frequency was picked to synchronize with the PETRA IV synchrotron radiation pulses. The gun provides a 15 mA average current at 300 kV. 
The bunches are 1.5 ns wide (FWHM) with rise and fall times of 300 ps. These characteristics produce a 2.4 nC bunch with peak current of 1.6 A. A 2-cm$^2$ gridded thermionic cathode provides the needed current \cite{LBL}. 

Both bunching cavities use quarter-wavelength coaxial cavities in the fundamental TEM mode, where the center coaxis is the beam pipe with a gap at one end. The lengths of the first and second bunchers are 123 and 44 cm, respectively. To ensure good transport into the accelerator structure, a solenoid was added. It will run at a few hundred gauss.  
Simulations show a low-energy tail on the bunch that must be removed before further acceleration. A chicane with a high-power energy slit capable of scrapping up to 30 kW, or 25 percent of the beam is needed between the injector and the accelerator. The high-power slit is approximately 1 meter long and the power density will be well below the 500W/cm$^2$ limit. Local x-ray shielding will be provided, but the 6 MeV beam will produce no neutrons and no component activation.  Downstream of the slit, the energy spread is 180 keV. The bunch is 33 ps long with a peak current of 74 A, a bunch charge of 2 nC, and a normalized rms emittance of 9 mm-mrad. The beam energy is 6.4 MeV.   It is possible to meet all of the design parameters with this approach \footnote{A number of choices of injector is available. Since the RF frequency of the ALS facility (499.65 MHz) is very close to the RF frequency of the PETRA storage ring (499.66 MHz), the design of the injector adopted in this document uses LBL injector design \cite{LBL}. An alternative approach that was considered would make use of an RF cavity with an internal photocathode.}.            

Each of the two acceleration section provides an energy gain of up to about 10 MeV. The 6-MeV beam from the injector is thus accelerated to a maximum of about 26 MeV on the first path and can then be recirculated and further accelerated to a maximum of about 46 MeV on the second path. The beam orbits must be separated at the exit of the second acceleration section and directed into their respective beam transport systems. Likewise, the beam combiner at the entrance to the first acceleration section must merge the orbits of the 6-MeV beam from the injector with that of the beam from the recirculating loop. Having passed through the accelerating structure, the 26 MeV electron beam is thus bent by 180 degrees in the horizontal plane. After it is used in the first FEL, the beam returns to the accelerating structure in the decelerating phase. In this configuration, the facility operates as a one-turn energy recovery Linac (ERL). To operate with the short wavelength FEL, the second 180-degree turn must be switched on.
The electron energy in the second FEL is 46 MeV. The beam is then decelerated two times and arrives  at the beam dump with low (6 MeV) energy.

\section{FEL Systems}

Depending on the re-circulation the maximum finite energy of electrons is 26 or 46 MeV. The first free electron laser includes one 4-m long electromagnetic undulator with a period of 12 cm and an optical cavity. The wavelength tuning range of this FEL is 30 - 100 mkm.  The optical cavity is 24,6 m long. Optical cavity mirrors are made of gold-platted copper, and water-cooled. The radiation is out-coupled through the holes in the mirror centers. The optical beamline is separated from the vacuum chamber by a diamond window installed at the Brewster angle.
The second FEL uses a variable gap permanent magnet undulator with period of 6 cm.  The wavelength tuning range of this FEL will amount to 5-30 mkm. The undulator is composed of a 30-period section. The radiation of both FELs is directed to the nitrogen-filled beamline to the user stations.   
%The basic optical cavity layout for second FEL is shown in Fig. 

\section{Optical Beam Transport}

Incorporating an infrared FEL source at the PETRA IV facility would necessitate propagating the infrared radiation from the FEL facility location to the PETRA IV experimental hall of the user's end stations. Since infrared beams are prone to significant diffraction, a suitable beam transport system must be provided to guide the beam for distances up to a few hundred meters, while maintaining it at a reasonable size. Moreover, the infrared beamline
should be designed to obtain a large transmission efficiency for radiation over a wide wavelength range. Usually,
the focusing of the middle and far infrared beam can only be achieved with reflective optics because lenses made from any material would reflect and absorb all radiation. To guide infrared radiation one can consider using mirror-based propagation lines or quasi-optical waveguides. The first approach, using mirror optics, is optimal for complex geometry where one needs to guide radiation around corners. In contrast, quasi-optical waveguides are naturally suited for long straight propagation lines.   In this design, similarly to in \cite{IRIS}, which focused on the LCLS baseline, it is proposed to use an open beam waveguide such as an iris guide, that is made of periodically spaced metallic screens with holes, for transporting the infrared and far infrared beam to the PETRA IV beam stations. In \cite{IRIS} it is already presented a complete iris guide theory, In particular, the requirements on the accuracy of the iris alignment were studied. 

The FEL radiation is extracted from the optical cavity through the out-coupling hole and enters the beamline passing a 1 mm thick diamond output window. Drayed nitrogen is circulated inside the beamline to avoid the absorption of infrared radiation by water vapor. After transportation, the FEL beam arrives through a polypropyleen window at one of the user stations in the PETRA IV experimental hall. It is possible to meet all design requirements with this approach \cite{K}.

\section{Acknowledgments}

I want to thank Ruediger Onken and Marcus Guehr for their valuable discussions. 
It is important to acknowledge that a significant portion of this work is based on the CDRL design report \cite{LBL}.
I am also thankful to Wim Leemans, who was closely involved in CDRL design at LBL, for his support and interest during the compilation of this work.

%\newpage

\end{document}